\documentclass[12pt]{article}
\usepackage{cite}
\usepackage{epsf}
\usepackage{graphicx}
\usepackage{psfrag}
\usepackage[dvips]{epsfig}
\usepackage[english]{babel}
\usepackage{amssymb,indentfirst}

\usepackage{amsmath}
\usepackage{verbatim}

\makeatletter
\renewcommand \thesection {\@arabic\c@section.}
\renewcommand\thesubsection   {\thesection\@arabic\c@subsection.}
\renewcommand\thesubsubsection{\thesubsection\@arabic\c@subsubsection.}
\makeatother

\def\starup#1{\mbox{$\raise1.8ex\hbox{$*$} \kern-.7em#1$}}
\def\krup#1{\mbox{$\raise1.8ex\hbox{$+$} \kern-1.0em#1$}}
\def\linup#1{\mbox{$\raise1.9ex\hbox{---} \kern-1.0em#1$}}

\begin{document}
\title{Resonance contribution of scalar color octet  \\
to $t \bar{t}$ production at the LHC % \\
in the minimal \\
four color quark-lepton symmetry model}
%\author{M.V.~Martynov\footnote{E-mail: martmix@mail.ru}, \,
% A.D.~Smirnov\footnote{E-mail: asmirnov@univ.uniyar.ac.ru}
\author{I.~V.~Frolov\footnote{E-mail: pytnik89@mail.ru},
M.~V.~Martynov\footnote{E-mail: martmix@mail.ru},
A.~D.~Smirnov\footnote{E-mail: asmirnov@uniyar.ac.ru}
\\
 \\
{\small Division of Theoretical Physics, Department of Physics,}\\
{\small Yaroslavl State University, Sovietskaya 14,}\\
{\small 150000 Yaroslavl, Russia.}}
%\address{}%
%\email{}%
\date{}
\maketitle
%\thanks{}%

%\vspase
% ----------------------------------------------------------------

\begin{abstract}
\noindent
The scalar color octet contribution to the resonance $t\bar{t}$-pair
production at the LHC is calculated and analysed with account of
the one loop effective two gluon vertex.
It is shown that this contribution from the scalar color octet $F_2$
predicted by the minimal model with the four color quark-lepton symmetry 
is for $\sqrt{s}=13$~TeV of about a few percents for $750 < m_{F_2} < 1800$ GeV
and can exceed 10\% for $400 < m_{F_2} < 750$~GeV. 
It is also pointed out that the search for the scalar octet~$F_2$
as the resonance in the dijet mass spectra seems to be difficult
because of the smallness of its one loop effective two gluon interaction.

\vspace{5mm}


\end{abstract}

The search for new physics effects beyond the Standard Model (SM)
is now one of the goals of the experiments at the LHC.
There are many models predicting new physics effects.
The most interesting of them look the models predicting new effects
due to enlarging the symmetry of the SM
because the search for such effects could help us to find the next symmetry
in yet unknown hierarchy of the symmetries which possibly unify
the known in the SM electroweak and strong interactions of quarks and leptons.

One of such variant of new physics beyond the SM can be induced
by the possible four color symmetry between quarks and leptons
of Pati-Salam type \cite{pati_salam}.
The four color quark-lepton symmetry can
be unified with the electroweak $SU_L(2) \! \times \! U(1) $ symmetry
of the SM in the minimal way by the group
\begin{equation}
\label{eq:GMQLS}
G_{MQLS}=SU_{V}(4)\times SU_{L}(2)\times U_{R}(1),
\end{equation}
%
%(the minimal quark-lepton-symmetric model - MQLS-model
%\cite{smirnov-1995-346,smirnov-yadphys-1995}),
%(MQLS-model, \cite{AD1,AD2})
\noindent
where the first factor is the vector-like group of the four color quark-lepton symmetry
the second one is the usual SM electroweak symmetry group for the left-handed fermions
and the third one is the corresponding hypercharge factor for the right-handed fermions
(the minimal quark-lepton symmetry model -- MQLS-model
\cite{smirnov-1995-346,AD22}).
As a remarkable consequence the four color quark-lepton symmetry predicts for
the electric charges of quarks and leptons the simple expression
\begin{equation}
\label{eq:Qemf}
Q^{em}_{f} \, = \, \sqrt{2/3} \,\, t_{15} \, + \,  \tau_3^L/2 + \, Y^R/2 \,
\nonumber
\end{equation}
in terms of the generators $t_{15}$, $\tau_3^L/2$
of the group~(\ref{eq:GMQLS}) with the hypercharge $Y^R=\pm 1$
for the ''up''  and ''down'' right-handed fermions,
which naturally explains the fractionary charges of quarks in terms
of the elementary charge.

As a result of the Higgs mechanism of splitting the masses of
quarks and leptons the four color symmetry in its minimal
realization on the gauge group~(\ref{eq:GMQLS})
predicts in addition to the SM Higgs doublet $\Phi^{(SM)}$ the existence
of the new scalar $SU_{L}(2)$-doublets
\begin{eqnarray}
%(15.2.1): \quad \,\,\,\,\,\,\,
\left ( \begin{array}{c} \Phi_{1}^{\,\prime}  \\
         \Phi_{2}^{\,\prime}\end{array} \right );
\left ( \begin{array}{c} S_{1\alpha}^{(+)}  \\
         S_{2\alpha}^{(+)}\end{array} \right );
\left ( \begin{array}{c} S_{1\alpha}^{(-)}  \\
         S_{2\alpha}^{(-)}\end{array} \right );
\left ( \begin{array}{c} F_{1c}  \\
         F_{2c}\end{array} \right )
%\left ( \begin{array}{c} \Phi_{1,15}^{(3)}  \\
%         \Phi_{1,15}^{(3)}\end{array} \right ),
\label{eq:SU2doublets}
\end{eqnarray}
with electric charges
\begin{eqnarray}
Q^{em}_{\Phi}:\quad \,\,\,\,\,\,\,
\left ( \begin{array}{c} 1  \\
         0 \end{array} \right );
\left ( \begin{array}{c} 5/3  \\
         2/3\end{array} \right );
\left ( \begin{array}{c} 1/3  \\
        - 2/3\end{array} \right );
\left ( \begin{array}{c} 1  \\
         0 \end{array} \right )
%\left ( \begin{array}{c} 1  \\
%         0 \end{array} \right )
\label{eq:Qem}
\nonumber
\end{eqnarray}
respectively.
The fields (\ref{eq:SU2doublets}) belong to the (1.2.1)+(15.2.1) representation
of the group (\ref{eq:GMQLS}), here the fields
$\Phi_{1}^{\,\prime}, \, \Phi_{2}^{\,\prime}$
form an additional colorless scalar doublet,
$S^{(\pm)}_{1\alpha}, \, S^{(\pm)}_{2\alpha},  \alpha=1,2,3$
are the color triplets forming two scalar leptoquark doublets and the fields
$F_{1c}, \, F_{2c}, c=1,2...8$ form
the scalar doublet of the color octets (the scalar gluon doublet).

Because of their Higgs origin the coupling constants of the
doublets (\ref{eq:SU2doublets}) with the fermions occur to be proportional
to the ratios $m_{f}/ \eta $ of the fermion masses $m_f$ to the SM
VEV~$\eta$ and are small for $u$-, $d$-, $s$-quarks
%($ m_u/ \eta \sim m_d/ \eta \sim 10^{-5},  m_s/ \eta \sim 10^{-3}$)
are more significant for $c$-, $ b$-quarks
%( $ m_c/ \eta \sim m_b/ \eta \sim 10^{-2}$)
and are especially significant for $t$-quark
( $ m_t/ \eta \sim 0.7$).
%(The more details concerning the interactions of the scalar
%doublets (\ref{eq:SU2doublets})
%with quarks and leptons can be found in~\cite{popov-2005-20}.)
As a result the scalar doublets~(\ref{eq:SU2doublets}) can manifest themselves
more probably in the processes with $t$-quarks.
In particular the scalar octet $F_2$ could manifest itself as
a resonance in $t\bar{t}$-pair production at the LHC.
It should be noted that the coupling constants of the
doublets (\ref{eq:SU2doublets}) with $t$-quark
are known (up to the mixing parameters), which gives the possibility
to estimate quantitatively the possible effects from these particles
in dependence on their masses.
The pair production of the scalar octets in $pp$-collisions at the LHC
has been discussed
in Refs \cite{manohar-2006-74, GrWise, Gerbush, Zerwekh, Perez, Drees,
Martynov:2008wf,martynov-smirnov-2010eng,GoncalvesNetto:2012nt, 
IdilbiPRD2010, CalvetJHEP2013, KuboPRL2014,Cortona}.

In the present paper we calculate the contribution of the scalar octet to
the cross section of the resonance $t\bar{t}$-pair production in pp-collisions
and analyse the possibility of manifestation of the scalar gluon $F_2$
of the MQLS-model as the corresponding resonance peak in $t\bar{t}$-pair
production at the LHC.

The details of interactions of the scalar doublets (\ref{eq:SU2doublets})
with quarks and leptons can be found in~\cite{popov-2005-20,raspady-yadphys-2007en}.
%
%The more details concerning the interactions of the scalar
%doublets (\ref{eq:SU2doublets})
%with quarks and leptons can be found in~\cite{popov-2005-20}.
%
%\cite{popov-2005-20},
%MQLS-model \cite{Smirnov:1995jq,AD22,popov-2005-20,Martynov:2008wf}
%
%xxxxx
%
In particular the interaction of the scalar gluon $F_{2}$
with up- and down-quarks in the MQLS-model has the chiral form
and can be written as
\begin{eqnarray}
%L_{F_1 u_i d_j} &=& \bar u_{i\alpha }  \Big [ ( h^L_{F_1})_{ij}P_L +
%(h^R_{F_1})_{ij}P_R  \Big ] (t_k)_{\alpha\beta}d_{j\beta} F_{1k} + {\rm h.c.},
%\nonumber\\
L_{ F_2 u_i u_j} &=& \bar u_{i\alpha} \Big [ ( h^L_{1F_2})_{ij}P_L
\Big ](t_c)_{\alpha\beta} u_{j\beta } F_{2c} + {\rm h.c.},
\label{eq:LF2uu}\\
L_{ F_2 d_i d_j} &=& \bar d_{i\alpha} \Big [ ( h^R_{2F_2})_{ij}P_R
\Big ](t_c)_{\alpha\beta} d_{j\beta } F_{2c} + {\rm h.c.},
\label{eq:LF2dd}
%\nonumber
\end{eqnarray}
%\newpage
where
$t_c, c=1,2\dots 8$ are the generators
of the $SU_c(3)$ group, $P_{L,R}=(1\pm\gamma_5)/2$ are the left and right
projection operators and
$(h^{L}_{1F_2})_{ij}, (h^{R}_{2F_2})_{ij}$ are the Yukawa coupling constants,
$i, j = 1, 2, 3$ are the generation indices.
As a result of the Higgs mechanism of generating the quark and lepton masses
the Yukawa coupling constants $(h^{L}_{1F_2})_{ij}, (h^{R}_{2F_2})_{ij}$
are defined by the expressions
\begin{eqnarray}
%(h^{L}_{F_1})_{ij} &=&\sqrt{ 3} \frac{1}{\eta \sin\beta}
%\Big [  m_{u_i} (C_Q)_{ij} -  (K^R_1)_{ik}m_{\nu_k}(\krup{K_1^L}C_l)_{kj} \Big ],
%\nonumber \\
%(h^{R}_{F_1})_{ij} &=&-\sqrt{ 3} \frac{1}{\eta \sin\beta}
%\Big [ (C_Q)_{ij}m_{d_i} -  (C_lK^L_2)_{ik}m_{l_k}(\krup{K_2^R})_{kj} \Big ],
%\nonumber \\
(h^{L}_{1F_2})_{ij} &=&-\sqrt{ 3} \frac{1}{\eta \sin\beta}
\Big [  m_{u_i} (\delta)_{ij} -  (K^R_1)_{ik}m_{\nu_k}(\krup{K_1^L})_{kj} \Big ],
%\nonumber
\label{eq:hL1F2ij}
\\
(h^{R}_{2F_2})_{ij} &=&-\sqrt{ 3} \frac{1}{\eta \sin\beta}
\Big [ m_{d_i} (\delta)_{ij} - (K^L_2)_{ik}m_{l_k}(\krup{K_2^R})_{kj} \Big ], 
%\nonumber
\label{eq:hR2F2ij}
\end{eqnarray}
where
$m_{u_i}, m_{d_i}$ and  $m_{\nu_k}, m_{l_k}$ are the masses
of up- and down-quarks and of neutrinos and charged leptons,
$K^{L,R}_1, K^{L,R}_2$
are the mixing matrices in leptoquark currents which are specific
for the model with the four color quark-lepton symmetry and
$\beta$ is a mixing angle of two colorless scalar doublets of MQLS model.
Among the coupling constants~(\ref{eq:hL1F2ij}),~(\ref{eq:hR2F2ij})
the largest is the constant $(h^{L}_{1F_2})_{33}$
which with neglect of the neutrinos masses takes the form
\begin{eqnarray}
\label{eq:hL1F233}
(h^{L}_{1F_2})_{33} &=&-\sqrt{ 3} \frac{m_t}{\eta \sin\beta}.
\end{eqnarray}
The interaction of the scalar gluon $F_{2}$ with $t$-quark
can be written as
\begin{equation}
\label{eq:LF2tt}
  L_{F_2 t t}= \bar{t}_\alpha(h_{F_2 t \bar t}^{S}+
h_{F_2 t \bar t}^{P} \,\gamma_5)(t_c)_{\alpha\beta} t_\beta F_{2c}+ {\rm h.c.},
\end{equation}
where 
%$(F_2)_{\alpha\beta} = (t_c)_{\alpha\beta} F_{2c}$ and 
the correspondent scalar and pseudoscalar coupling constants
with account of~(\ref{eq:hL1F233}) take the form
\begin{eqnarray}
\label{eq:hphsF2tt}
h^{S}_{F_2t\bar{t}} =h^{P}_{F_2t\bar{t}}=
-\frac{\sqrt{3}}{2} \frac{m_t}{\eta \sin\beta} % \approx -\frac{0.61}{\sin\beta}
\approx - 0.61/ \sin\beta.
\end{eqnarray}
The coupling constants~(\ref{eq:hphsF2tt}) increase with decreasing $\sin\beta$
so that for $\sin\beta= 1, \, 0.7, \, 0.4$ the perturbation theory parameters
take the values
$(h^{S,P}_{F_2t\bar{t}})^2/4 \pi \approx  0.03, \, 0.06, \,  0.18$ respectively.
Below we restrict ourselves by the mixing angle region
$0.4\leq\sin\beta\leq 1$.

The interactions~(\ref{eq:LF2uu}),~(\ref{eq:LF2dd}) lead to the decays
$F_2 \to u_{i} \bar u_{i}, \, F_2 \to d_{i} \bar d_{i}$ and in the case of
$m_{F_2} > 2 m_t$ the decay $F_2 \to t \bar t$ is dominant with
the width~\cite{popov-2005-20,raspady-yadphys-2007en}
\begin{eqnarray}
\label{eq:GF2tt}
\Gamma(F_2\to t\bar t)&=&
m_{F_2}\frac{3}{32\pi}\left(\frac{m_t}{\eta}\right)^2
\left(1-2\frac{m_t^2}{m_{F_2}^2}\right)
\sqrt{1-4\frac{m_t^2}{m_{F_2}^2}}
\frac{1}{\sin^2\beta}.
%\label{eq:5e}
\end{eqnarray}
For the masses $m_{F_2}=400- 2000$ GeV the width~(\ref{eq:GF2tt}) is
of about $(2-30)/\sin^2\beta$ GeV
and $\Gamma_{F_2}/m_{F_2}=(0.5-1.5)\%/\sin^2\beta$.

As seen from the expressions~(\ref{eq:hL1F2ij}),~(\ref{eq:hR2F2ij})
the coupling constants of the interaction of the scalar gluon $F_{2}$
with $u$- and  $d$- quarks  are of order of
$ m_u/ \eta \sim m_d/ \eta \sim 10^{-5}$
and the interactions of these quarks as the initial partons with
the scalar gluon~$F_{2}$ are negligibly small. On the other hand
the Lagrangian \eqref{eq:LF2tt}, \eqref{eq:hphsF2tt} can induce
through the loop contribution of $t$-quark
the more significant effective interaction of two initial gluons
with the scalar gluon~$F_{2}$, which should be taken into account.
The analogous effective two gluon interaction is induced also with
the colorless scalar $\Phi_{2}^{\,\prime}$.

The calculation of the effective two gluon vertex of interaction with the scalar octet
is like to that with the colorless scalar and we perform below these calculations
simultaneously. For this purpose we write the flavour diagonal interactions
of scalar octet and of the scalar color singlet with quarks in the model
independent form as
\begin{equation}
\label{LPhiffbar}
  L_{\Phi q \bar q}= \bar{q}_\alpha(h_{\Phi q \bar q}^{S}+
h_{\Phi q \bar q}^{P}\gamma_5)\Phi_{\alpha\beta} q_\beta + {\rm h.c.},
\end{equation}
where  $\Phi_{\alpha\beta}=%C^{\Phi_0q\bar{q}}_{\alpha\beta}\Phi_0=
\Phi_0\delta_{\alpha\beta}$ for the colorless scalar particle $\Phi_0$ and
$\Phi_{\alpha\beta}=
%C^{\Phi_8q\bar{q}}_{k\alpha\beta}\Phi_{8k}=
\Phi_{8c}(t_c)_{\alpha\beta}$ for the scalar octet $\Phi_8$,
$t_c$  are the generators of the $SU_c(3)$ group ($c=1,2...8$),
$h_{\Phi q \bar q}^{S}$ and  $h_{\Phi q \bar q}^{P}$ are the corresponding
scalar and pseudoscalar coupling constants.
For the MQLS-model the scalars $\Phi_8$ and $\Phi_0$  correspond to
$F_{2}$ and $\Phi_{2}^{\,\prime}$ respectively.

\begin{figure}[htb]
\begin{center}
\includegraphics[scale=0.5,keepaspectratio=true]{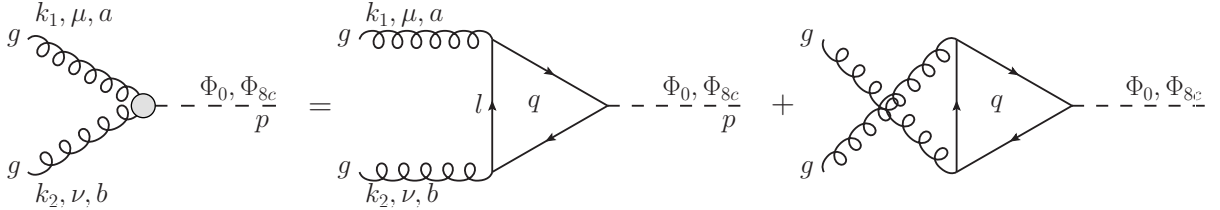}
\caption{The effective vertex $\Gamma^{(q)\mu\nu}_{ab\Phi}(p,k_1,k_2)$
induced by one loop contribution of quark~$q$ for $\Phi = \Phi_0, \Phi_{8c}$.}
\label{Veff_diag}
\end{center}
\end{figure}

The effective vertex $\Gamma^{(q)\mu\nu}_{ab\Phi}(p,k_1,k_2)$
of interaction of two gluons with scalar field $\Phi = \Phi_0, \Phi_{8c}$
induced by the Lagrangian \eqref{LPhiffbar}
with account of one loop contribution of quark~$q$ is described
by the diagrams in the Fig. \ref{Veff_diag} 
\newline
and can be written as
%
%\begin{align}
\begin{eqnarray}
\label{Gexplicit}
\nonumber
&& \hspace{-17mm} \Gamma^{(q)\mu\nu}_{ab\Phi}(p,k_1,k_2)= \\
\nonumber
&& \hspace{-14mm}  = c_{ab\Phi}^{(1)} \, g_s^2
\int\frac{d^n l}{i(2\pi)^n}\frac{\mathrm{Tr}\big((h_{\Phi qq}^S+h_{\Phi qq}^P\gamma^5)
(\hat{l}
+\hat{k}_1+m_q)\gamma^\mu(\hat{l}+m_q)\gamma^\nu(\hat{l}-\hat{k}_2+m_q) \big) }
{\left((l+k_1)^2-m_q^2+i\varepsilon\right)
\left(l^2-m_q^2+i\varepsilon\right)
\left((l-k_2)^2-m_q^2+i\varepsilon\right)} + \\
&& \hspace{-14mm}  + \, c_{ab\Phi}^{(2)} \, g_s^2
\int\frac{d^n l}{i(2\pi)^n}\frac{\mathrm{Tr}\big((h_{\Phi qq}^S+h_{\Phi qq}^P\gamma^5)
(\hat{l}+\hat{k}_2+m_q)\gamma^\nu(\hat{l}+m_q)\gamma^\mu(\hat{l}-\hat{k}_1+m_q) \big) }
{\left((l+k_2)^2-m_q^2+i\varepsilon\right)
\left(l^2-m_q^2+i\varepsilon\right)
\left((l-k_1)^2-m_q^2+i\varepsilon\right)},
\end{eqnarray}
%\end{align}
%
where $c_{ab\Phi}^{(1)}, \, c_{ab\Phi}^{(2)}$ are the color factors,
$c_{ab\Phi_0}^{(1)}=c_{ab\Phi_0}^{(2)}=\delta^{ab}/2$
for the colorless particle $\Phi_0$ and
$c_{ab\Phi_{8c}}^{(1)}=\mathrm{Tr}(t_a t_b t_c)=\frac{1}{4}(d_{abc}+if_{abc})$,
$c_{ab\Phi_{8c}}^{(2)}=\mathrm{Tr}(t_b t_a t_c)=\frac{1}{4}(d_{abc}-if_{abc})$
for the color octet $\Phi_8$, $a, b, c = 1, 2, ..., 8$ are the color indices,
 $d_{abc}$ and $f_{abc}$  are the $d$- and $f$- constants of the $SU_c(3)$ group.

To regulate ultraviolet divergences in one loop calculation of
$\Gamma^{(q)\mu\nu}_{ab\Phi}(p,k_1,k_2)$
we use the dimensional regularization  with $n = 4-2\varepsilon$
and use for $\gamma_5$  the Larin's prescription~\cite{Larin:1993tq} based
on the t'Hooft-Veltman scheme \cite{THooft1972}.

With account of the contributions of all the quarks the resulted effective vertex
$\Gamma^{\mu\nu}_{ab\Phi}(p,k_1,k_2)$
in the case of real gluons ($k_1^2=0$, $k_2^2=0$, $\hat{s}=p^2=2(k_1 k_2)$)
can be parametrized as
\begin{eqnarray}
&&  \hspace{-12mm} \Gamma^{\mu\nu}_{ab\Phi}(p,k_1,k_2)  =
\sum_q  \Gamma^{(q)\mu\nu}_{ab\Phi}(p,k_1,k_2) =
\nonumber
\\
&& \hspace{-12mm}  =-C_{ab\Phi} \frac{\alpha_s \sqrt{\hat{s}}}{\pi}\left[
\left(g^{\mu\nu}-\frac{2k_1^\nu k_2^\mu}{\hat{s}}\right)F^{S}_{\Phi }(\hat{s})
-2 i \varepsilon^{\mu\nu\rho\sigma}\frac{k_{1 \rho} k_{2 \sigma}}{\hat{s}} \,
F^{P}_{\Phi }(\hat{s})+\frac{2k_1^\mu k_2^\nu}{\hat{s}} \,\,
G^{S}_{\Phi}(\hat{s})\right]
\label{eq:GFgg}
\end{eqnarray}
by the form factors
\begin{eqnarray}
\label{F1Fgg}
&& \hspace{-12mm} F^{S,P}_{\Phi}(\hat{s})=
\sum_q h_{\Phi q \bar q}^{S,P}\,\tilde{F}^{S,P}(\hat{s},m_q^2), \hspace{12mm}
G^{S}_{\Phi}(\hat{s})=\sum_q h_{\Phi q \bar q}^{S}\,\tilde{G}^{S}(\hat{s},m_q^2),
%\\
%&& \hspace{-12mm} \tilde{F}^{S}(\rho_q)=
%\frac{m_q}{\sqrt{\hat{s}}}\left[(\hat{s}-4m_q^2) \, C_0(0,0,\hat{s},m_q^2,m_q^2,m_q^2)-
%2\right],\\
%&& \hspace{-12mm} \tilde{F}^{P}(\rho_q)=m_q\sqrt{\hat{s}} \,\, C_0(0,0,\hat{s},m_q^2,m_q^2,m_q^2),\\
%&& \hspace{-12mm} \tilde{G}^{S}(\hat{s},m_q^2)=
%\frac{m_q}{\sqrt{\hat{s}}}\left[(\hat{s}+4m_q^2) \, C_0(0,0,\hat{s},m_q^2,m_q^2,m_q^2)+
%4B_0(\hat{s},m_q^2,m_q^2)-\frac{4 A_0(m_q^2)}{m_q^2}\right],
\end{eqnarray}
where $C_{ab\Phi} $ is the color factor with
 \begin{equation}
%C_{\Phi_0}= C_{\Phi_0}^{ab}  =
C_{ab\Phi_0} = \delta_{ab}/2 \equiv C_{ab}  , \,\,  \,\,
%C_{\Phi_8}=C_{\Phi_8}^{abc}=
C_{ab\Phi_{8c}} =  d_{abc}/4 \equiv C_{abc}
 \end{equation}
for the color singlet $\Phi_0$ and for the color octet $\Phi_8$.

For the form factors $\tilde{F}^{S,P}(\hat{s},m_q^2), \, \tilde{G}^{S}(\hat{s},m_q^2)$
we have found the expressions
\begin{eqnarray}
%\label{F1Fgg}
%&& \hspace{-12mm} F^{S,P}_{\Phi}(\hat{s})=
%\sum_q h_{\Phi q \bar q}^{S,P}\,\tilde{F}^{S,P}(\rho_q), \hspace{12mm}
%G^{S}_{\Phi}(\hat{s})=\sum_q h_{\Phi q \bar q}^{S}\,\tilde{G}^{S}(\hat{s},m_q^2),
%\\
\label{eq:FSt}
&& \hspace{-12mm} \tilde{F}^{S}(\hat{s},m_q^2)=
\frac{m_q}{\sqrt{\hat{s}}}\left[(\hat{s}-4m_q^2) \, C_0(0,0,\hat{s},m_q^2,m_q^2,m_q^2)-
2\right]  \equiv \tilde{F}^{S}(\rho_q)   ,\\
\label{eq:FPt}
&& \hspace{-12mm} \tilde{F}^{P}(\hat{s},m_q^2)=m_q\sqrt{\hat{s}} \,\,
C_0(0,0,\hat{s},m_q^2,m_q^2,m_q^2) \equiv \tilde{F}^{P}(\rho_q)   ,\\
\label{eq:GSt}
&& \hspace{-12mm} \tilde{G}^{S}(\hat{s},m_q^2)=
\nonumber
\\ &&  \hspace{-12mm}
 = \frac{m_q}{\sqrt{\hat{s}}}
\bigg[
%\left[
(\hat{s}+4m_q^2) \, C_0(0,0,\hat{s},m_q^2,m_q^2,m_q^2)+
%\\ &&
4B_0(\hat{s},m_q^2,m_q^2) -
%\\ &&
\frac{4 A_0(m_q^2)}{m_q^2}
%\right]
\bigg ]
\equiv \tilde{G}^{S}(\rho_q),
\end{eqnarray}
where
$A_0$, $B_0$, $C_0$ are the Passarino--Veltman (PV)
scalar integrals \cite{Passarino:1978jh,Denner:1991kt}, % and
with account of the explicit form of these integrals
the form factors~(\ref{eq:FSt})--(\ref{eq:GSt})
depend only on the variable  $\rho_q = \sqrt{\hat{s}}/m_q $.
The form factors~(\ref{eq:FSt})--(\ref{eq:GSt}) 
account the one loop contribution of quark $q$ 
in the model independent way,  
the specific features of the model are presented %only 
by the coupling constants  $h_{\Phi q \bar q}^{S,P}$ 
in the form factors~(\ref{F1Fgg}).

\begin{figure}[htb]
\begin{center}
\begin{tabular}{cc}
  %\includegraphics[scale=0.8]{fff} & \includegraphics[scale=0.8]{F3ReIm}\\
   \includegraphics[scale=0.4]{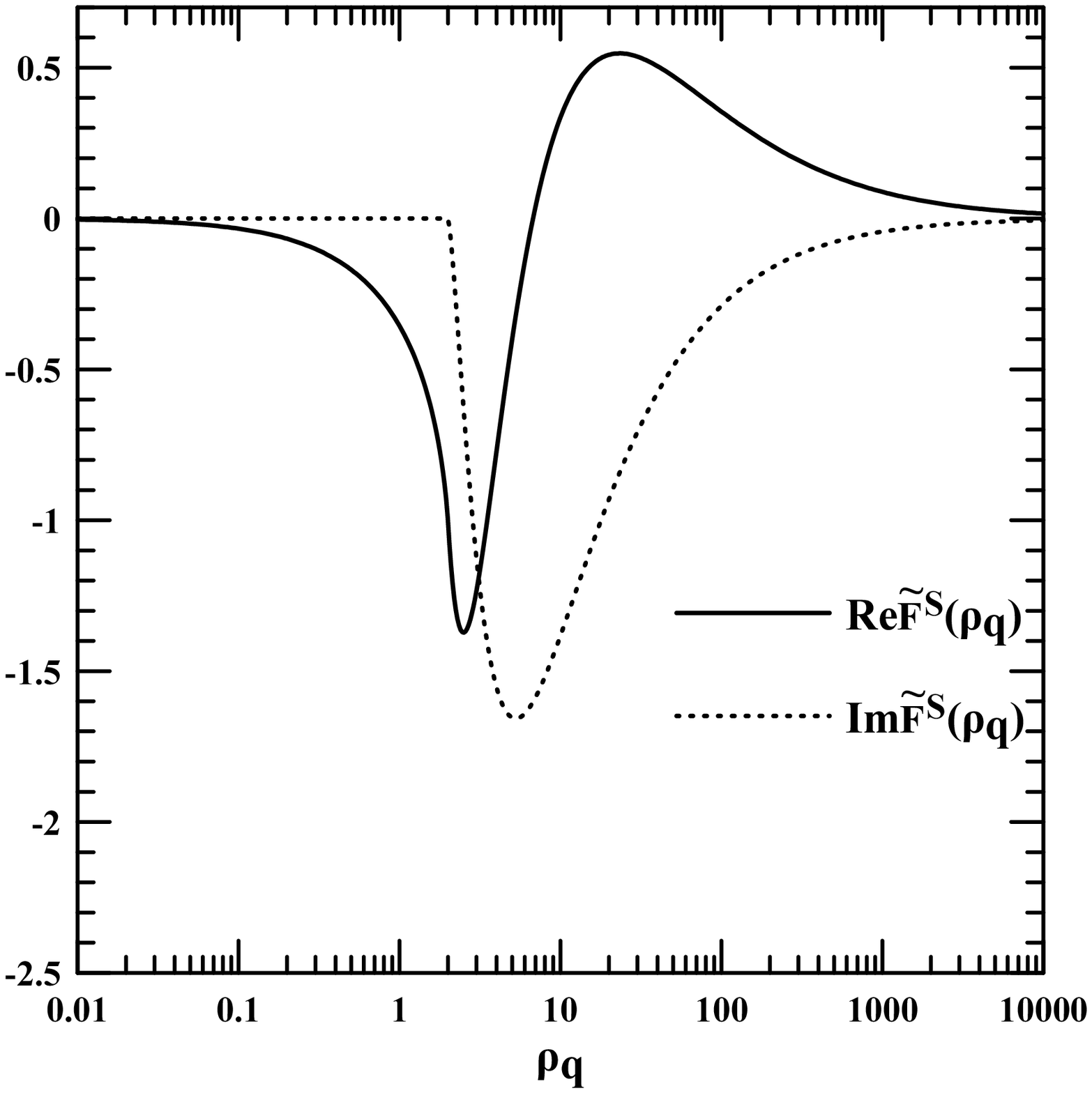} & \includegraphics[scale=0.4]{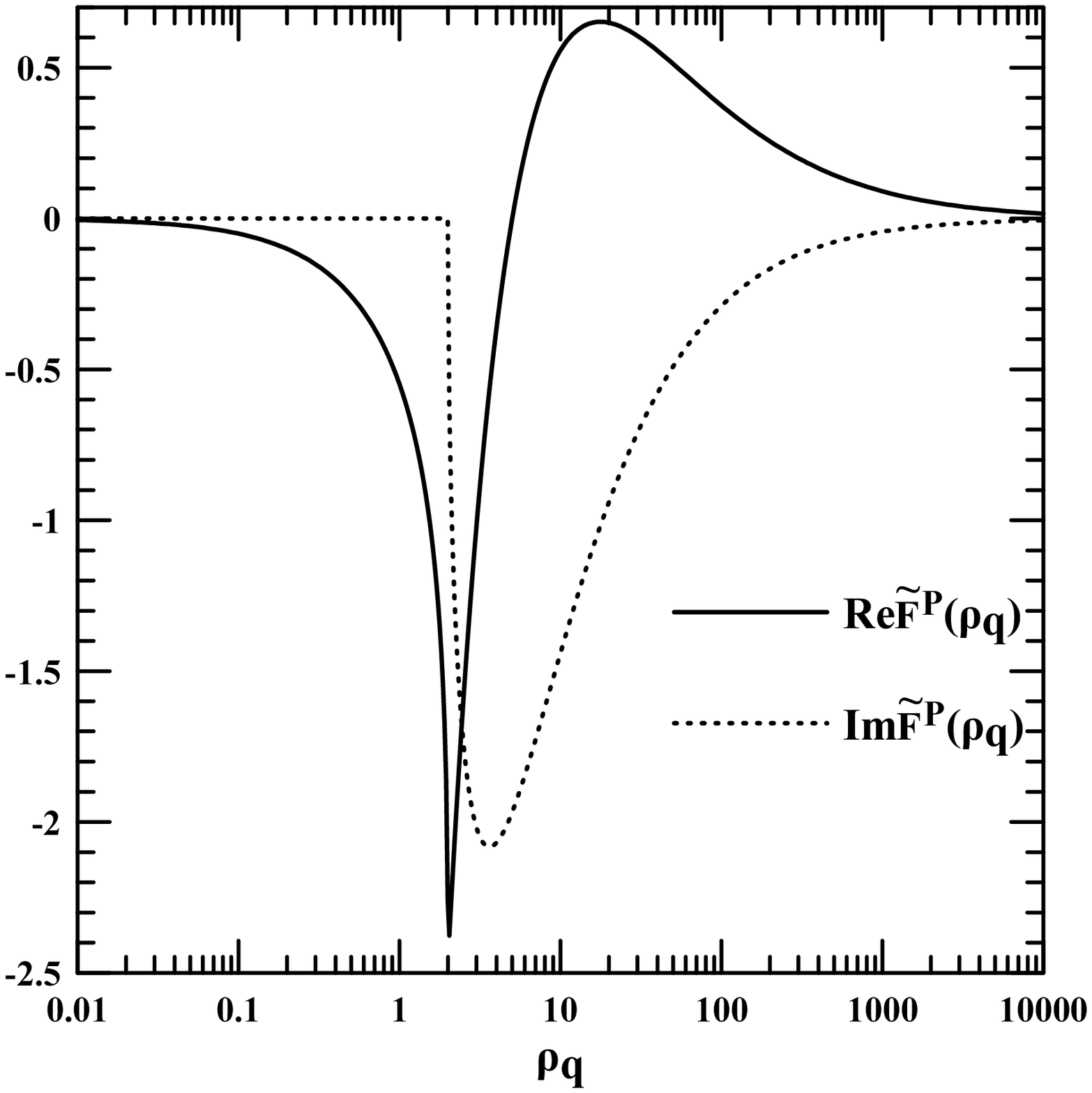}\\
a) & b)
\end{tabular}
\end{center}
\caption{The real and imaginary parts of form factors a)  $\tilde{F}^{S}(\rho_q)$ and
b) $\tilde{F}^{P}(\rho_q)$ as the~function of $\rho_q=\sqrt{\hat{s}}/m_q$. }
\label{Ftildes}
\end{figure}

Due to the gauge invariance the longitudinal component
of the vertex~(\ref{eq:GFgg}) %which is
parametrized by the form factor $G^{S}_{\Phi}(\hat{s})$ do not enter
to the observed variables.
The real and imaginary parts of the form factors
$\tilde{F}^{S}(\rho_q),\tilde{F}^{P}(\rho_q)$
as a functions of $\rho_q$ are shown in the Fig.~\ref{Ftildes}.
%\newpage
%

\begin{figure}[htb]
\begin{center}
\includegraphics[scale=0.55,keepaspectratio=true]{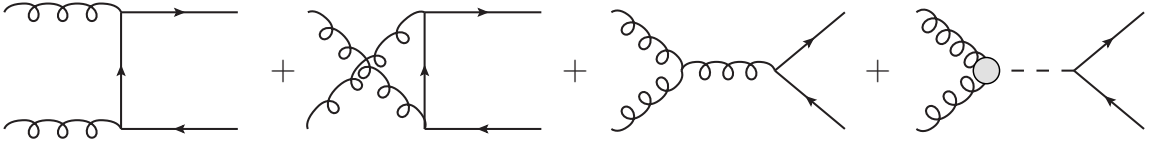}
\caption{Diagrams of the process $gg\to Q \bar{Q}$ in the SM LO and with account of
the effective vertex $\Gamma^{\mu\nu}_{ab\Phi}(p,k_1,k_2)$.
}
\label{ggttVeffHF_diag}
\end{center}
\end{figure}

We have calculated the cross section of the process $gg \rightarrow  Q \bar{Q}$
of $Q\bar{Q}$ pair production in gluon fusion in QCD LO with account also
of the effective vertex \eqref{eq:GFgg}.
The diagrams of this process are shown in the Fig. \ref{ggttVeffHF_diag}.

The total cross section of the process $gg \rightarrow  Q \bar{Q}$ can be written as
the sum
\begin{equation}
\label{Sectggttgen}
\sigma_0(gg{\rightarrow}  Q \bar{Q}, \mu)=
\sigma^{\mathrm{SM}}_{0}(gg{\rightarrow} Q \bar{Q}, \mu)+
\Delta\sigma^{\Phi}(gg {\rightarrow} Q \bar{Q}, \mu)
\end{equation}
of the well known QCD LO cross section
\begin{eqnarray}
\label{SectggttExplic1}
%\nonumber
\sigma^{\mathrm{SM}}_{0}(gg{\rightarrow} Q \bar{Q}, \mu)
&=&\frac{\alpha_s^2(\mu) \, \pi}{48\hat{s}}
\left[\left(
 v^4-18 v^2+33
\right)\log{\frac{1+ v}{1- v}}+ v(31 v^2-59)
\right],
\end{eqnarray}
where $ v=\sqrt{1-4m_{Q}^{2}/\hat{s}}$ is the velocity of quark $Q$ in the center
of mass frame, $\hat{s}$ is the squared energy in the center of momentum
frame of the gluons, $\mu$  is a typical mass scale of the process,
and the contribution
$\Delta\sigma^{\Phi}(gg {\rightarrow} Q \bar{Q}, \mu)$
to this process from the scalar $\Phi$.

For the contribution $\Delta\sigma^{\Phi}(gg {\rightarrow} Q \bar{Q}, \mu)$
we have found the expression
\begin{eqnarray}
\label{SectggttExplic2}
\nonumber
&& \hspace{-8mm} \Delta\sigma^{\Phi}(gg {\rightarrow} Q \bar{Q}, \mu)=
\\
\nonumber
&& \hspace{-8mm}
=\hspace{-1mm}\frac{\tilde{C}_{\Phi}^{(1)}}{64}\frac{\alpha_s^2(\mu) m_Q}
{\pi\sqrt{\hat{s}}}\frac{\mathrm{Re}\left[
\left(\hat{s}-m_\Phi^2-i m_\Phi\Gamma_\Phi\right)\left(-(h_{\Phi Q \bar Q}^{S})^* v^2F^{S}_{\Phi}(\hat{s})-
(h_{\Phi Q \bar Q}^{P})^*F^{P}_{\Phi}(\hat{s})\right)
\right]
}
{(\hat{s}-m_\Phi^2)^2+m_\Phi^2\Gamma_\Phi^2}
\log{\frac{1\hspace{-1mm}+\hspace{-1mm}v}{1\hspace{-1mm}-\hspace{-1mm}v}}+
\\                                                    
\label{SectggttExplic3}
&& \hspace{-8mm}
+2\frac{\tilde{C}_{\Phi}^{(2)}}{2048}\frac{\alpha_s^2(\mu) v \hat{s}}{\pi^3}
\frac{|h_{\Phi Q \bar Q}^{S}|^2 v^2+|h_{\Phi Q \bar Q}^{P}|^2}{(\hat{s}-m_\Phi^2)^2+
m_\Phi^2\Gamma_\Phi^2}\left(
|F^{S}_{\Phi}(\hat{s})|^2+|F^{P}_{\Phi}(\hat{s})|^2
\right),
\end{eqnarray}
where the form factors  $F^{S,P}_{\Phi}(\hat{s})$
are given by the expressions~(\ref{F1Fgg}),(\ref{eq:FSt}),(\ref{eq:FPt})
and $\tilde{C}_{\Phi}^{(1)}, \, \tilde{C}_{\Phi}^{(2)}  $ are the color factors
with
\begin{eqnarray}
\label{eq:Ct0}
&& \hspace{-40mm} \tilde{C}_{\Phi_0}^{(1)} = C_{ab} C_{ab} = 2,
\hspace{15mm} \tilde{C}_{\Phi_0}^{(2)}= C_{ab} C_{ab} \, n_c = 6, \,\,\,
\\
\label{eq:Ct8}
&& \hspace{-40mm} \tilde{C}_{\Phi_8}^{(1)}= C_{abc} C_{abc} = 5/6 ,
\hspace{8mm} \tilde{C}_{\Phi_8}^{(2)}= C_{abc} C_{abc}/2 = 5/12
\end{eqnarray}
for the color singlet $\Phi_0$ and for the color octet $\Phi_8$ respectively,
$n_c$ is the number of colors of the $SU_c(n_c)$ group, the numerical values
in~(\ref{eq:Ct0}),~(\ref{eq:Ct8}) correspond to the $SU_c(3)$ group.

The total cross section $\sigma_{tot}(p p \rightarrow t \bar{t})$
of the $t\bar{t}$ production in $pp$-collisions with account of the contribution
of scalar octet $F_2$ can be written as the sum
\begin{eqnarray}
\label{FullSect}
\sigma_{tot}(p p \rightarrow t \bar{t}) =
\sigma^{\mathrm{SM}}(pp{\rightarrow} t \bar{t})+
\Delta\sigma^{F_2}(pp {\rightarrow} t \bar{t})
\end{eqnarray}
of the SM cross section $\sigma^{\mathrm{SM}}(pp{\rightarrow} t \bar{t})$
and the contribution $\Delta\sigma^{F_2}(pp {\rightarrow} t \bar{t})$
induced by scalar gluon $F_2$ via effective 
vertex \eqref{eq:GFgg}--\eqref{eq:FPt}.

We obtain the total cross section from partonic cross 
sections~\eqref{SectggttExplic1}, \eqref{SectggttExplic3} 
by integrating the expression
\begin{align}
\label{dspp1x1x2}
\nonumber
\hspace{-2mm} \frac{d \sigma_{tot}(pp  \rightarrow  t \bar{t})}{ dx_1 dx_2} =
\sum_k F_k^{p\bar{p}}(x_1,x_2,\mu_f) K(s)  
\sigma^{SM}_0(\bar{q_k} q_k {\rightarrow}  t \bar{t},\mu) +
\\ +
F_g^{pp}(x_1,x_2,\mu_f)
%\big(
K(s)\sigma^{SM}_0(gg  {\rightarrow} t \bar{t},\mu) +
%\\ +
F_g^{pp}(x_1,x_2,\mu_f)
\Delta\sigma^{F_2}(gg {\rightarrow}  t \bar{t},\mu)
%\big)
%dx_1 dx_2
\end{align}
over the variables $ 0 \le  x_1, x_2 \le 1 $,
where  $ x_1, x_2 $ are partonic parts of the momenta of protons,
$\hat{s} = x_1 x_2 s$,
$ \;s = (P_1+P_2)^2$, \, $P_1, P_2$ are the momenta of the colliding protons,
$\sigma^{SM}_0(\bar{q_k} q_k {\rightarrow}  t \bar{t},\mu)$ is 
the well known SM LO cross section,
$K(s)$ is the $K$-factor, which we use for the better agreement of the
SM LO predictions of the cross section of  $t\bar{t}$-pair production
with the corresponding aNNNLO SM predictions~\cite{Kidonakis:2014isa}.
The partonic functions in~\eqref{dspp1x1x2} are given as
\begin{eqnarray}
\label{Fkpp1}
F_k^{pp}(x_1,x_2,\mu_f) &=& f_{q_k}^{p}(x_1,\mu_f) f_{\bar{q}_k}^{p}(x_2,\mu_f) +
f_{\bar{q}_k}^{p}(x_1,\mu_f) f_{q_k}^{p}(x_2,\mu_f),
%\nonumber
\\
F_g^{pp}(x_1,x_2,\mu_f) &=& f_{g}^{p}(x_1,\mu_f) f_{g}^{p}(x_2,\mu_f),
\label{Fgpp1}
%\nonumber
\end{eqnarray}
where
$f_{q_k}^{p}(x,\mu_f)$, $f_{\bar{q}_k}^{p}(x,\mu_f)$,
$f_{g}^{p}(x,\mu_f)$ are the parton distribution functions of quark~$q_k$
of flavor~$k$, antiquark $\bar{q}_k$ and gluons in the proton,
$\mu_f$ is the factorization scale.

For numerical calculations we use the analytical expressions for scalar PV
integrals $A_0$, $B_0$, $C_0$ from the Denner's paper~\cite{Denner:1991kt} and
we also perform the cross check with using
LoopTools/FF~\cite{Hahn:1998yk,vanOldenborgh:1990yc}.

We have calculated the cross section~\eqref{FullSect}
at $\sqrt{s} = 7, 8, 13$~TeV with using the parton distribution functions
MSTW2008 \cite{Martin:2009iq} (NNLO, $\mu=\mu_f=m_t$, $m_t=173.21$ GeV).
For calculations we use the values of
$K$-factors $K(s)  =1.7110$, $1.7095$, $1.7024$
for energies $\sqrt{s} = 7, 8, 13$~TeV respectively,
in this case the cross section
$\sigma^{\mathrm{SM}}(pp{\rightarrow} t \bar{t})$
reproduces well the aNNNLO SM predictions for the cross section
of $t\bar{t}$ production~\cite{Kidonakis:2014isa}.

We have calculated the contributions $\Delta\sigma^{F_2}(pp {\rightarrow} t \bar{t})$
to the total cross section of the $t \bar{t}$ production from the scalar gluon $F_2$
defined by the last term in the expression~\eqref{dspp1x1x2}
for $\sqrt{s} = 7, 8, 13$~TeV, $m_{F_2} = 400 \div 1000$~GeV
and $\sin\beta= 1, 0.7, 0.4$.
These contributions for %the parameters values  
%$\sin\beta= 0.4$,  $m_{F_2} \lesssim 460$~GeV 
%
\begin{equation}
%\begin{eqnarray}
\sin \beta = 0.4, \, \, \, \,  m_{F_2}  \lesssim 460 \, \, \mbox{GeV} 
%\nonumber
\label{ATLASCMSlimftcs}
%\end{eqnarray}
\end{equation}
exceed the experimental unsertainties of the measurements 
of the total $t\bar{t}$ cross sections performed 
by ATLAS($20.3 fb^{-1}$)~\cite{AadEPJC2016}  
and CMS($19.7 fb^{-1}$)~\cite{KhachatryanJHEP2016}    
at $\sqrt{s}=7, 8$~TeV   
and by ATLAS($3.2 fb^{-1}$)~\cite{AaboudPLB2016}  
%and CMS($43 pb^{-1}$)~\cite{KhachatryanPRL2016_1}        
and CMS($2.3 fb^{-1}$)~\cite{CMS_PAS_TOP_2016}        
at $\sqrt{s}=13$~TeV      
(hence the values~\eqref{ATLASCMSlimftcs} are excluded by these data)   
whereas the other parameters regions are consistent with these data within 
the experimental unsertainties. 
As seen the exclusion limit~\eqref{ATLASCMSlimftcs}  
resulting from the data on the total $t\bar{t}$ cross sections is low.      

\begin{figure}[htb]
\begin{center}
\includegraphics[scale=0.5,keepaspectratio=true]{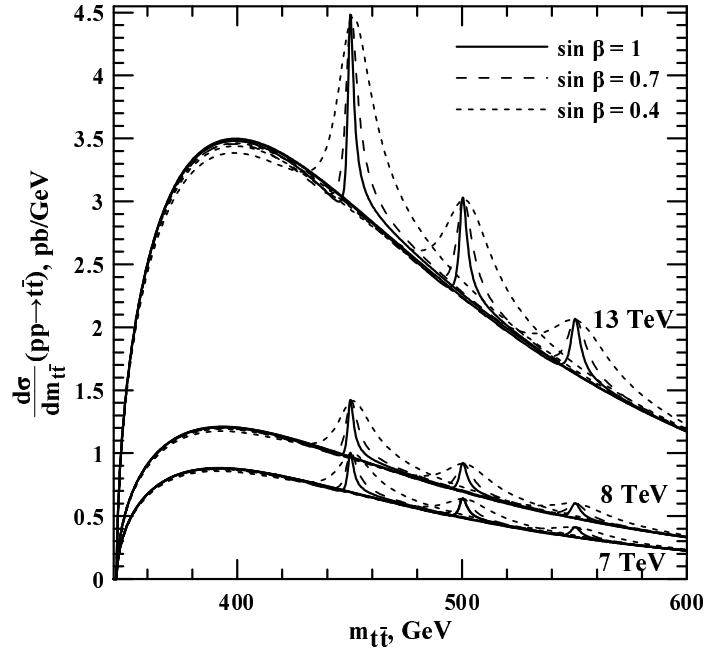}
\caption{The invariant mass spectrum $d\sigma(pp  \to  t\bar{t})/dm_{t\bar{t}}$
of the $t\bar{t}$-pair production in $pp$ collisions
at the LHC at energies $\sqrt{s}=7, 8, 13$ TeV with account of the contributions
of scalar gluon $F_2$ with masses $m_{F_2}=450, 500, 550$ GeV
for $\sin \beta=1, 0.7, 0.4$.}
\label{dsmtt}
\end{center}
\end{figure}

Using the known relations between the variables  $ x_1, x_2 $ and
the invariant mass $m_{t\bar{t}}$ of $t\bar{t}$ pair and
the rapidity $y$ of the final $t$-quark
\begin{equation}
\label{peremennie}
    m^2_{t\bar{t}} = x_1 x_2 s, \;\; y=\ln{\frac{x_1}{x_2}}, \;\;\;\;\;\;\;\;
x_{1,2} = \frac{m_{t\bar{t}}}{\sqrt{s}} \, e^{\pm y/2}
\end{equation}
and integrating the expression~\eqref{dspp1x1x2} over the rapidity $y$
with account of the scalar gluon contribution
$\Delta\sigma^{F_2}(gg {\rightarrow}  t \bar{t},\mu_f)$
we obtain the invariant mass
spectrum $d\sigma(pp  \to  t\bar{t})/dm_{t\bar{t}}$ in the form
\begin{align}
\label{dsppMtt}
\hspace{-2mm}
\frac{d \sigma_{tot}(pp  \rightarrow  t \bar{t})}{ d m_{t\bar{t}} } =
\hspace{-1mm}
\frac{m_{t\bar{t}}}{s} \int_{-\ln{(s/m_{t\bar{t}}^2})}^{+\ln{(s/m_{t\bar{t}}^2})}
\frac {d \sigma_{tot}(pp  \rightarrow  t \bar{t}) }{ dx_1 dx_2 } \, dy.
\end{align}

In the same way but with neglect of the scalar gluon contribution
$\Delta\sigma^{F_2}(gg {\rightarrow}  t \bar{t},\mu_f)$ we obtain
the background invariant mass spectrum
$d\sigma_b(pp  \to  t\bar{t})/dm_{t\bar{t}}$ which is in agreement
with the theoretical predictions \cite{Ahrens:2010zv} and with the experimental
results \cite{Aad:2014zka,Chatrchyan:2012saa,Khachatryan:2015oqa}.

The invariant mass spectra $d\sigma(pp  \to  t\bar{t})/dm_{t\bar{t}}$
at the LHC energies $\sqrt{s}=7,8,13$ TeV
for the scalar gluons masses $m_{F_2}=450, \,500, \, 550$ GeV
and for $\sin \beta=1, \, 0.7, \, 0.4$
are shown in the Fig.~\ref{dsmtt}.
As seen in the Fig.~\ref{dsmtt} the distribution~\eqref{dsppMtt}
%slightly exceeds the background spectrum and
at $m_{t\bar{t}} \sim  m_{F_2}$ 
has the typical peaks induced by the scalar gluon $F_2$ with widths depending 
on $\sin \beta$.

To distinguish the signal and background events we use the significance
%estimator~\mbox{\cite{Bartsch:824351,BITYUKOV1998}}
estimator~\mbox{\hspace{-0.8mm}\cite{Bartsch:824351,BITYUKOV1998}}
\begin{equation}
\label{Significance}
%    \mathcal{S}=\sqrt{2 \left((N_s+N_b)\ln{\left(1+\frac{N_s}{N_b}\right)}-N_s\right)},
%    \mathcal{S}=\sqrt{ 2  \big[ \, (N_s+N_b)\ln{\left(1+N_s/N_b \right)}-N_s \, \big]} \, ,
%    \mathcal{S}=\sqrt{ 2  \big[ \, (n_s+n_b)\ln{\left(1+n_s/n_b \right)}-n_s\,\big]} \, ,
\mathcal{S}=2(\sqrt{ n_s+n_b}-\sqrt{n_b}),
\end{equation}
where $n_s$ ($n_b$) -- are number of signal (background) events in
the $t\bar{t}$ invariant  mass region $m \pm \Delta m / 2$
near the scalar gluon mass $m_{F_2}$.
These numbers can be calculated as
\begin{eqnarray}
  n_s &=& n_{tot}-n_{b}, \\
  n_{tot,\,b} &=& L \, \sigma_{tot,\,b}(m,\Delta m),
%\\
\label{eq:ntotnb}
\end{eqnarray}
\begin{equation}
\sigma_{tot,\,b}(m,\Delta m)=
\int_{m-\frac{1}{2}\Delta m}^{m+\frac{1}{2}\Delta m}
\frac{d\sigma_{tot,\,b}(pp \to t\bar{t})}{dm_{t\bar{t}}} dm_{t\bar{t}},
\end{equation}
where $L$ is integrated luminosity.
\begin{table}[htb]
  \centering
  \caption{
The signal significance $\mathcal{S}$ and the ratio $n_s/n_b$
in the bin ranges $m \pm \Delta m / 2$
of CMS data on $t\bar{t}$ production
at $\sqrt{s}=8$~TeV~\cite{Khachatryan:2015oqa} for $\sin \beta=1(0.4)$ 
}
  \begin{tabular}{|c|c|c||c|c|c|c|}
       \hline
$m_{t\bar{t}}$ bin range [GeV] &  $\delta_{\mathrm{Syst}}$  &  $\delta_{\mathrm{Total}}$ 
& { $m_{F_2}$ }  & $\sin \beta$ & $\mathcal{S}$ & $n_s/n_b$  \\
  $\lbrace m \pm \Delta m / 2\rbrace $    &   [\%] &  [\%]& [GeV] &  &  & [\%] \\
      \hline
      $345-400$ $\lbrace 372.5\pm22.5\rbrace $    &  7.1 &   7.5 & 373 & 1(0.4) & 59(327)&5.8(33.2) \\
             &           &       & 400 & 1(0.4)& 11(18) & 1.0(1.7)\\
\hline
      $400-470$ $\lbrace 435\pm35\rbrace $   &  2.9 &   3.6 & 435 & 
      1(0.4)& 32(174) &2.7(14.8) \\
             &           &       & 470 & 
             1(0.4)& 3.5(21) & 0.2(1.7)\\
\hline
      $470-550$ $\lbrace 510\pm40\rbrace $   &  6.1 &   6.4 & 510 & 1(0.4) & 22(113) &2.1(11.3) \\
             &           &       & 550 & 1(0.4)& 0.6(26) &0.05(2.5) \\
      \hline
      $550-650$ $\lbrace 600\pm50\rbrace $   &  7.3 &   7.7 & 600 & 1(0.4) & 13(68) &1.7(8.5) \\
             &           &       & 650 & 1(0.4)& 1.8(20) &0.2(2.5) \\
      \hline
      $650-800$ $\lbrace 725\pm75\rbrace $   &  4.2 &   4.9 & 725 & 1(0.4) & 6.7(35) &1.0(5.5) \\
             &           &       & 800 & 1(0.4)& 1.3(11) &0.2(1.7) \\
      \hline
     \end{tabular}
\label{Tab1}
\end{table}

The signal significance $\mathcal{S}$ and the ratio $n_s/n_b$
in the bin ranges $m \pm \Delta m / 2$ of CMS data on $t\bar{t}$ production
at $\sqrt{s}=8$~TeV~\cite{Khachatryan:2015oqa}
are shown for corresponding masses $m_{F_2}$ and for $\sin \beta=1(0.4)$
in the Table~\ref{Tab1},  $\delta_{\mathrm{Syst}}$ and $\delta_{\mathrm{Total}}$
are the experimental systematic and total relative uncertainties
of the values of the cross sections in the bin ranges.
As seen from the Table~\ref{Tab1} for 
\begin{equation}
%\begin{eqnarray}
\sin \beta = 0.4, \, \, \, \,  m_{F_2}< 725 \, \, \mbox{GeV} 
%\nonumber
\label{CMSlimfdcs}
%\end{eqnarray}
\end{equation}
the ratios $n_s/n_b$ exceed the experimental relative 
uncertainties $\delta_{\mathrm{Total}}$ and hence the values~\eqref{CMSlimfdcs}  
are excluded by the data of ref.~\cite{Khachatryan:2015oqa}. 
At the same time for $\sin \beta=1$ and for all the masses $m_{F_2}$  
the ratios $n_s/n_b$ do not exceed the experimental relative uncertainties 
and the scalar gluon $F_2$ in this case can not be visible 
in the data of ref.~\cite{Khachatryan:2015oqa}.
The exclusion limit~\eqref{CMSlimfdcs} resulting from the current CMS data on 
the differential $t\bar{t}$ cross section at $\sqrt{s}=8$~TeV is slightly higher 
then the limit~\eqref{ATLASCMSlimftcs} nevertheless it is also rather low.      

In the case of $\sqrt{s}=13$~TeV we have calculated the ratios $n_s/n_b$
in dependence on  $m_{F_2}$ and $\sin \beta$ assuming $\Delta m=100$~GeV
(as close to the experimental facilities) and
to maximize the significance estimator $\mathcal{S}$
we use also the optimized bin width $\Delta m = 1.28 \, \Gamma_{F_2}$ which 
corresponds to the $3 \sigma$ width in the case of a Gaussian distribution.

\begin{figure}[htb]
%\center{\includegraphics[scale=0.5]{nSnBLHC13_2}}
\center{\includegraphics[scale=0.5]{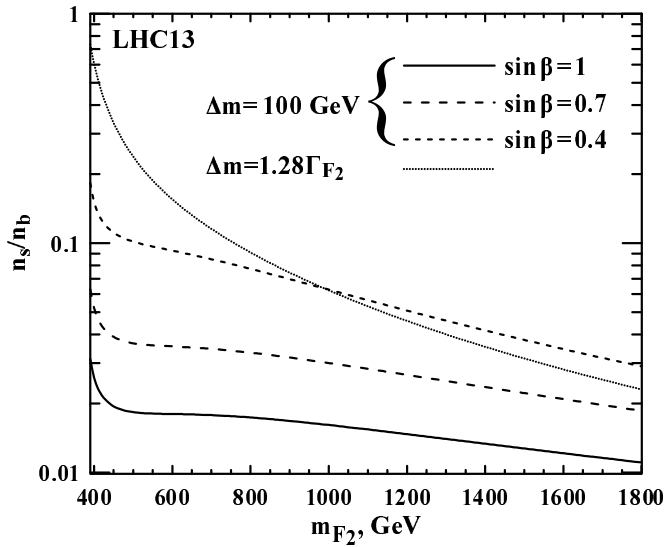}}
\caption{ The ratios $n_s/n_b$ for the $t\bar{t}$ production at the LHC
with account of the scalar gluon~$F_2$ contribution
%at the LHC
for $\sqrt{s}=13$~TeV
and bin sizes $\Delta m=100$ GeV and $\Delta m=1.28 \, \Gamma_{F_2}$
 as the functions of $m_{F_2}$.
%for $\sin \beta=1, 0.7, 0.4$.
}
\label{nSnB13}
\end{figure}

For $\sqrt{s}=13$~TeV the ratios $n_s/n_b$ as the functions
of $m_{F_2}$ for $\sin \beta=1, 0.7, 0.4$ are shown in the Fig.~\ref{nSnB13}
(in the case of $\Delta m = 1.28 \, \Gamma_{F_2}$
the ratio $n_s/n_b$ practically does not depend on $\sin \beta$).
As seen from the Fig.~\ref{nSnB13} for $ 0.4 \lesssim \sin \beta \le 1 $  
the ratios $n_s/n_b$ in the region $500 < m_{F_2} < 1800$ GeV 
for $\Delta m=100$~GeV and in the region $750 < m_{F_2} < 1800$~GeV  
for $\Delta m = 1.28 \, \Gamma_{F_2}$ 
are of about a few percents (do not exceed 10\%).   
By this reason the search for such scalar octet~$F_2$
as the resonance peak in $t\bar{t}$-pair production at the LHC 
in these regions can need the experimental relative accuracy 
in measuring the corresponding cross section of about one percent.
In the region $400 < m_{F_2} < 750$ GeV for $\Delta m = 1.28 \, \Gamma_{F_2}$ 
the ratio $n_s/n_b$ exceeds 10\% and in this case 
the experimental relative accuracy of about a few percent 
can be sufficient to search for the $F_2$ resonance peak 
in $t\bar{t}$-pair production with using the optimized bin width.   
It should be noted that the current total (systematic) experimental accuracy 
in measuring the differential $t\bar{t}$ cross section at $\sqrt{s}=8$~TeV 
in the range $400 < m_{t \bar{t}} < 1600$ GeV is of about 
$\delta_{\mathrm{Total}} (\delta_{\mathrm{Syst}}) =(3.6(2.9) - 12.7(9.8))\%$
~\cite{Khachatryan:2015oqa}. 
If one assumes that the experimental accuracy in measuring the differential 
$t\bar{t}$ cross section at the initial stage of the experiments 
at $\sqrt{s}=13$~TeV  will be of the same order as that for $\sqrt{s}=8$~TeV 
the search for the $F_2$ resonance peak in $t\bar{t}$-pair production 
in the region $750 < m_{F_2} < 1800$ GeV at this stage 
will be rather difficult but will be possible in 
the region $400 < m_{F_2} < 750$ GeV with using the optimized bin width.   
At the present time at $\sqrt{s}=13$~TeV  there are measured the total 
$t\bar{t}$ cross sections by ATLAS($3.2 fb^{-1}$)~\cite{AaboudPLB2016}   
%and CMS($43 pb^{-1}$)~\cite{KhachatryanPRL2016_1} 
and CMS($2.3 fb^{-1}$)~\cite{CMS_PAS_TOP_2016} 
Collaborations with 
the accuracies of 4.4\% and 3.9\% respectively which are of the same order 
as the accuracies of 4.3\% and 3.7\% of the corresponding measurements 
of the total $t\bar{t}$ cross sections by 
ATLAS($20.3 fb^{-1}$)~\cite{AadEPJC2016}    
and CMS($19.7 fb^{-1}$)~\cite{KhachatryanJHEP2016} 
at $\sqrt{s}=8$~TeV.       

The recent CMS data~\cite{Khachatryan:2015dcf} on search for narrow resonances
in the dijet mass spectra at $\sqrt{s}=13$~TeV and  $L=2.4\, \mathrm{fb}^{-1}$
exclude in particular the color-octet scalar $S_8$
with the effective interaction~\cite{Han:2010rf,Chivukula:2014pma}
\begin{equation}
\label{eq:LS8}
\mathcal{L}_{S_8}=g_s d_{abc}\frac{k_S}{\Lambda_S} S^a_8 G^b_{\mu\nu}G^{c \, \mu\nu}
\end{equation}
for the masses $m_{S_8} < 3.1$~TeV with assuming $\Lambda_S=m_{S_8}$ and $k_S=1$.
In the effective Lagrangian~\eqref{eq:LS8}  $\Lambda_S$ is a typical mass scale
and $k_S$  is an arbitrary dimensionless parameter.
%It should be noted however that
The interaction~\eqref{eq:LS8}
in notation~\eqref{eq:GFgg} corresponds to the form factor
$F^{S}_{S_8}(\hat{s})=
16\pi^{3/2}\frac{k_S}{\sqrt{\alpha_s}}\frac{\sqrt{\hat{s}}}{\Lambda_S}$
which is real and for  $\Lambda_S=3100$~GeV and $k_S=1$
in the region $\sqrt{\hat{s}}=400 \div 4000$~GeV takes the large
values $F^{S}_{S_8}(\hat{s}) \approx 35 \div 347$.
At the same time the real part of the form factor $F^{S}_{F_2}(\hat{s})$ defined for
the scalar octet $F_2$  by the equations~\eqref{eq:GFgg}--\eqref{eq:FSt}
for $\sqrt{\hat{s}}=400 \div 4000$~GeV has the values of order unity,
$Re(F^{S}_{F_2}(\hat{s}))= - (- 0.79 \div 0.31)/\sin \beta$.
It means that the effective parameter $k_S$  in~\eqref{eq:LS8}
for the scalar octet $F_2$ with $m_{F_2}=3100$~GeV
in the region $\sqrt{\hat{s}}=400 \div 4000$ GeV has the values
$k_S^{F_2} = - (- 0.023 \div 0.0009)/\sin \beta$
which are essentially less in magnitude than the unity.
In the region $\sqrt{\hat{s}}=2500 \div 3500$~GeV including the mass $m_{F_2}=3.1$~TeV
the parameter $k_S^{F_2}$ has the very small in magnitude values
$k_S^{F_2} = - ( 0.0011 \div 0.0012)/\sin \beta $
and as a result the quoted lower exclusion mass limit $m_{S_8} < 3.1$~TeV 
becomes for the scalar octet $F_2$ essentially more lower.
Moreover keeping also in mind that the dominant decay of the scalar octet $F_2$
is the decay $F_2 \to t \bar t$ and hence $Br(F_2 \to jj) \ll 1$
it will be difficult to see the scalar octet $F_2$
as the resonance in the dijet mass spectra.

As seen, the search for the scalar octet $F_2$ as the resonance peak 
in $t\bar{t}$ production will be a some experimental problem needing the relatively 
high accuracy in measuring the differential $t\bar{t}$ cross section 
and is difficult in the case of the search for the corresponding resonance peak  
in the dijet mass spectra.  In this situation the other way to find 
the possible manifestation of the scalar octet $F_2$ in $pp$-collisions at the LHC
can be the process of the $F_2 \bar{F_2}$-pair production followed by the decays of 
the $F_2 \bar{F_2}$ pairs to $t\bar{t} t\bar{t}$ quarks, 
$pp \to F_2 \bar{F_2} \to t\bar{t} t\bar{t}$.   
The cross section of the pair production of the scalar octets in $pp$-collisions  
occurs to be rather large and can be accessible for the experimental investigations 
at the LHC 
\cite{manohar-2006-74, GrWise, Gerbush, Zerwekh, Perez, 
Drees, Martynov:2008wf,martynov-smirnov-2010eng,GoncalvesNetto:2012nt, 
IdilbiPRD2010, CalvetJHEP2013,KuboPRL2014%,Cortona
}.  
One should keep in mind however that in many models
\cite{manohar-2006-74, GrWise, Gerbush, Zerwekh, Perez, 
%Drees, Martynov:2008wf,martynov-smirnov-2010eng,GoncalvesNetto:2012nt, 
IdilbiPRD2010, CalvetJHEP2013%,KuboPRL2014,Cortona
}   
the interactions of the scalar octets with quarks have ambiguities 
in the coupling constants, which gives no possibility to estimate 
quantitatively the scalar octet contributions to the cross section 
of $t\bar{t} t\bar{t}$ production. 
Unlike this the scalar octet $F_2$ has the known coupling constants~(\ref{eq:hphsF2tt})   
of interaction with $t$ quark which are sufficiently large. 
As a result the scalar octet $F_2$ can give the contribution to the cross section 
of $t\bar{t} t\bar{t}$ production which can be measurable 
at the LHC \cite{Martynov:2008wf,martynov-smirnov-2010eng}.   
Keeping also in mind that the SM cross section of $t\bar{t} t\bar{t}$ production 
(which forms the background) is less then that of the $t\bar{t}$ production   
it would be interesting to search for the scalar octet $F_2$ also as the peak 
in the $t\bar{t}$ invariant mass spectra among the $t\bar{t} t\bar{t}$ events.  
  
In conclusion, we summarize the results of this paper.

The effective vertex of interaction of the scalar color octet with two gluons
is calculated with account of the one loop quark contribution.
With account of this interaction the contribution of the scalar color octet 
to the partonic cross section of resonance $Q\bar{Q}$-pair production 
in the gluon fusion is calculated.

The total and differential cross sections of the $t\bar{t}$ production
in $pp$-collisions at the LHC are calculated  with account
of the resonance contribution of scalar color octet~$F_2$ predicted by 
the minimal model with the four color quark-lepton symmetry 
and analysed in dependence on two parameters of the model,
the $F_2$ mass $m_{F_2}$ and mixing angle $\beta$. 

From the comparison with the CMS data on the differential cross sections 
of  $t\bar{t}$ production  at $\sqrt{s}=8$~TeV~\cite{Khachatryan:2015oqa} it is shown 
that the scalar color octet $F_2$ with $\sin \beta=0.4$, $m_{F_2}< 725$ GeV 
is excluded by these data but for $\sin \beta=1$ and for all the masses $m_{F_2}$ 
the scalar color octet $F_2$ gives the contribution to this process 
of about a few percents and can not be visible in these data.

For $\sqrt{s}=13$~TeV it is shown that the contribution
of the scalar color octet $F_2$ with $750 < m_{F_2} < 1800$ GeV to 
resonance $t\bar{t}$-pair production at the LHC is of about 
a few percents (do not exceed 10\%) and the search for such scalar color octet 
in this process in this mass region can need the experimental accuracy 
of about one percent in measuring the corresponding cross section, 
at the same time in the region $400 < m_{F_2} < 750$ GeV 
this contribution can exceed 10\% and in this case  
the experimental relative accuracy of about a few percent 
can be sufficient to search for the $F_2$ resonance peak 
in $t\bar{t}$-pair production with using the optimized bin width.

Taking the mass limits resulting from the recent CMS 
data~\cite{Khachatryan:2015dcf}
on the search for narrow resonances in the dijet mass spectra 
for the color-octet scalar $S_8$ into account it is pointed out that because
of the smallness of the effective two gluon coupling constant and of 
the dijet branching ratios $Br(F_2 \to jj) \ll 1$ the search for 
the scalar color octet~$F_2$ as the resonance in the dijet mass spectra 
seems to be difficult.

%\bibliographystyle{elsarticle-num_mod}
%\bibliographystyle{ws-mpla}
%\bibliography{D:/Book1/for_my_work/my_base1}

\end{document}